\documentclass[a4paper]{article}

\usepackage{INTERSPEECH2019}
\usepackage{hyperref}

\title{Training Multi-Speaker Neural Text-to-Speech Systems \\ using Speaker-Imbalanced Speech Corpora}
\name{Hieu-Thi Luong$^{1,2}$, Xin Wang$^1$, Junichi Yamagishi$^{1,2,4}$, Nobuyuki Nishizawa$^3$}
\address{
  $^1$National Institute of Informatics, Tokyo, Japan \\
  $^2$SOKENDAI (The Graduate University for Advanced Studies), Kanagawa, Japan \\
  $^3$KDDI Research Inc., Saitama, Japan
  $^4$University of Edinburgh, Edinburgh, UK
  }
\email{\{luonghieuthi,wangxin,jyamagis\}@nii.ac.jp, no-nishizawa@kddi-research.jp}

\begin{document}

\maketitle
\begin{abstract}
  When the available data of a target speaker is insufficient to train a high quality speaker-dependent neural text-to-speech (TTS) system, we can combine data from multiple speakers and train a multi-speaker TTS model instead. Many studies have shown that neural multi-speaker TTS model trained with a small amount data from multiple speakers combined can generate synthetic speech with better quality and stability than a speaker-dependent one. However when the amount of data from each speaker is highly unbalanced, the best approach to make use of the excessive data remains unknown. Our experiments showed that simply combining all available data from every speaker to train a multi-speaker model produces better than or at least similar performance to its speaker-dependent counterpart. Moreover by using an ensemble multi-speaker model, in which each subsystem is trained on a subset of available data, we can further improve the quality of the synthetic speech especially for underrepresented speakers whose training data is limited.

\end{abstract}

\section{Introduction}
Recent advances in statistical parametric speech synthesis research have produced synthetic speech indistinguishable from natural speech when a model is trained with a large and high quality speech corpus \cite{luong2018investigating,shen2017natural}.
However to scale the technology to multiple voices and reduce the production cost, the ability to build TTS systems from a smaller and less refined corpus is crucial.
As data sparsity is the major challenge for this task, many schemes have been proposed to alleviate it. If the speech corpus is created from scratch, the sentence corpus used for recording could be carefully designed to ensure a balanced coverage of linguistic units \cite{zhu2002corpus,bozkurt2003text}.
A less refined speech corpus, such as a corpus of found data, can also be used by filtering out utterances deemed unfit \cite{cooper2017utterance,lee2018comparison}.
A data selection scheme can also be applied on legacy corpora to remove redundant samples \cite{Podsiadło2018,kuo2018data}.
In another approach, we could combine speech data from many speakers and train a multi-speaker TTS system \cite{yamagishi2010thousands}.


Recent neural acoustic models are capable of achieving high performance for both single speaker modeling \cite{shen2017natural} and multi-speaker modeling \cite{jia2018transfer,chen2018sample} tasks. 
The multi-speaker model is simple to set up \cite{fan2015multi,zhao2016speaker} and can generate more stable speech waveforms than those of the speaker-dependent model when the amount of the target speaker's data is limited \cite{jia2018transfer}. Latorre et al.\ \cite{latorre2018effect} compared the performances of multi-speaker and single-speaker models using different amounts of data and reported similar results for various conditions. In these multi-speaker experiments \cite{chen2018sample,latorre2018effect}, the number of utterances contributed by each speaker is kept perfectly or roughly balanced.
In this paper, we are interested in finding the best strategy to train a multi-speaker model using an existing speaker-unbalanced corpus.


Class imbalance is a common issue faced by many classification systems because real-world data are usually predominated by the normal classes while lacking samples of the abnormal classes. Many techniques have been proposed to tackle this problem. Over-sampling and under-sampling are simple and effective approaches to obtain synthetically balanced corpus \cite{chawla2002smote}. In this paper, we use the same techniques to prepare the training set for a multi-speaker acoustic model. Moreover, we propose using an ensemble model, which combines predictions of multiple subsystems, to produce a better prediction itself. Our ensemble acoustic model for speech synthesis shares the same spirit as the ensemble deep learning system for speech recognition \cite{deng2014ensemble}.

In section \ref{sec:methodology} of this paper, we describe our methodology for multi-speaker acoustic and the ensemble models. Section \ref{sec:experiments} provides details about the experimental conditions and Section \ref{sec:evaluations} presents both objective and subjective evaluation results of our proposal. We conclude in Section \ref{sec:conclusions} with with a brief summary and mention of future work.


\section{Multi-speaker and ensemble models}
\label{sec:methodology}

\subsection{Multi-speaker model for speaker-imbalanced corpus}
In this paper we adopt the same auto-regressive neural-network acoustic model used in our prior publication \cite{luong2018investigating}. By appending a one-hot vector speaker code to every frame of the linguistic input $\boldsymbol{x}$, we created a multi-speaker model that can generate multiple voices simply by changing the speaker code.
The method is simple but effective and does not depend on the network architecture \cite{zhao2016speaker,hojo2018dnn}.
This essentially means that all parameters of the network are shared among all training speakers except the bias of the first hidden layer:
\begin{equation}
\label{eq:spkembedding}
    \boldsymbol{h}_{1} = \tanh( \boldsymbol{W}_{1}\boldsymbol{x} + \boldsymbol{c}_1 + \boldsymbol{b}^{(k)} )
\end{equation}
where $\boldsymbol{h}_{1}$ is the output of the first hidden layer containing $m$ units, $\boldsymbol{W}_1 \in \mathbb{R}^{m \times m}$ and $\boldsymbol{c}_1 \in \mathbb{R}^{m \times 1}$ are common parameters shared among all speakers, and $\boldsymbol{b}^{(k)} \in \mathbb{R}^{m \times 1}$ is a speaker-specific bias projected from the speaker's one-hot vector. $\tanh$ is the non-linear activation function of the first hidden layer.

As most of the network parameters are shared and stochastically trained with combined data, using an imbalanced corpus might produce a model that is over-trained on the majority speakers while under-trained on the minority.
To test this hypothesis we apply resampling techniques, which are widely used to create synthetically balanced datasets \cite{kubat1997addressing,chawla2002smote}. Here, we can choose to perform under-sampling \cite{kubat1997addressing} of the majority speakers, over-sampling of the minority speakers \cite{japkowicz2000class}, or a little of both \cite{chawla2002smote}. While these techniques are commonly used for classification tasks, we applied them in the context of training a multi-speaker neural acoustic model.

\begin{figure}[t]
  \centering
  \includegraphics[width=0.9\linewidth]{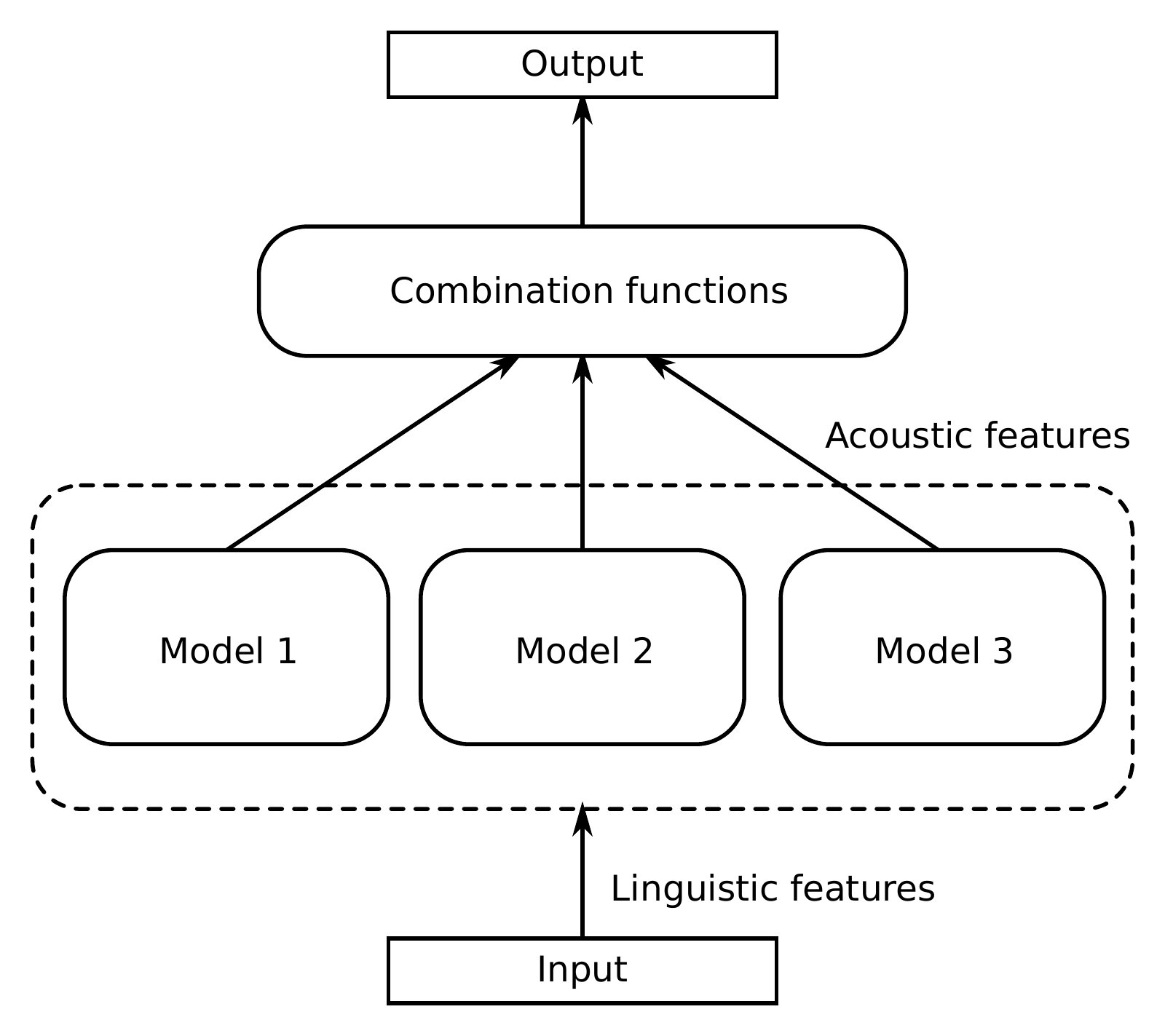}
  \vspace{-2mm}
  \caption{Ensemble multi-speaker acoustic model used for our investigation.}

  \label{fig:ensemble}
  \vspace{-3mm}
\end{figure}

\begin{table*}[t]
\caption{Data sets of target speakers.}
\label{tab:data}
\centering
\scalebox{1.0}{
  \begin{tabular}{lrrrrrrrrrr}
  \hline
  \hline
  \textbf{Speaker ID} & \textbf{XS01} & \textbf{XS02} & \textbf{S03} & \textbf{S04} & \textbf{S05} & \textbf{M06} & \textbf{M07} & \textbf{M08} & \textbf{L09} & \textbf{XL10} \\ \hline
  Training (unique utterances): \\
  \hspace{3mm} Speaker-Dependent & 735 & 994 & 1393 & 1568 & 1749 & 3024 & 3983 & 4364 & 5516 & 8750 \\ 
  \hspace{3mm} Sampling 1\textsuperscript{st} & 728 & 938 & 1227 & 1341 & 1444 & 1901 & 2088 & 2179 & 2320 & 2532 \\
  \hspace{3mm} Sampling 2\textsuperscript{nd} & 729 & 955 & 1214 & 1340 & 1442 & 1892 & 2074 & 2185 & 2312 & 2516 \\
  \hspace{3mm} Sampling 3\textsuperscript{rd} & 722 & 944 & 1242 & 1329 & 1418 & 1916 & 2122 & 2186 & 2325 & 2554 \\
  \hspace{3mm} Ensemble (Sampling 1+2+3) & 735 & 994 & 1391 & 1559 & 1742 & 2869 & 3541 & 3807 & 4424 & 5630 \\
    Validation & 50 & 50 & 50 & 50 & 50 & 50 & 50 & 50 & 50 & 50  \\ 
  Testing & 100 & 100 & 100 & 100 & 100 & 100 & 100 & 100 & 100 & 100 \\ \hline
  \end{tabular}
}
\vspace{-1mm}
\end{table*}

\subsection{Linear ensemble for acoustic feature inference}
\label{subsec:ensemble}
In addition to the resampling techniques, we also investigate using stacking \cite{wolpert1992stacked,breiman1996stacked} to combine the predictions of several systems in the hope of further reducing the mismatch between generated and real-life samples. Ensemble learning is a method of using multiple models to obtain a better performance; it is used in many other research fields \cite{ren2016ensemble}.
For example, Deng and Platt \cite{deng2014ensemble} performed a linear combination of the original speech-class posterior probabilities provided by subsystems at the frame level for automatic speech recognition (ASR). Their ensemble model capitalizes on the diversity of neural network architectures to provide diverse prediction outputs.

Our ensemble model, shown in Fig.\ref{fig:ensemble}, shares many traits with the model proposed in \cite{deng2014ensemble} for ASR. 
To create diverse subsystems, we used the same network architecture in each subsystem but trained them on different data subsets randomly sampled from a training corpus. This strategy is more straightforward than creating subsystems with varied network architectures \cite{deng2014ensemble,chebotar2016distilling}.
Moreover we take a much simpler and non-parametric approach for the combination functions to test our hypothesis.
Deterministic average-based combination functions are defined to combine the output of the subsystems. As the two main acoustic features used in our experiments are mel-generalized cepstral coefficients (MGCs) and fundamental frequency (F0), we define the combination functions as follows:
\begin{itemize}
    \item \textbf{Combination function for MGC}: As the MGCs at each frame are continuous values, our ensemble model simply computes the average of the MGCs produced by the subsystems.
    \item \textbf{Combination function for F0}: Because the F0 is a continuous value at a voiced frame but a discrete symbol (i.e., unvoiced flag) at an unvoiced frame, we first decide whether one frame is voiced or unvoiced by voting. If most of the subsystems generated voiced F0s values, we take the average F0 value as the ensemble model's output. Otherwise, the output F0 is set to unvoiced.
\end{itemize}

\section{Experiments}
\label{sec:experiments}

\subsection{Dataset and features}

Our experiments are data-driven and we seek to identify the best approach to train a speech synthesis system from an imbalanced speech corpus. The corpus we used contained utterances from ten female Japanese speakers, who are professional or at least familiar with voice acting work. The number of utterances of each speaker ranged from 1,000 to 10,000. After processing and removing utterances unsuitable for speech synthesis, we split the remaining data into training, validation and testing sets, as displayed in Table~\ref{tab:data}.
As we applied a sampling technique to create a synthetic speaker-balanced corpus, the number of unique utterances of each speaker obtained from these sampling sessions are also included in Table~\ref{tab:data}.

The acoustic features used in our experiments consist of 60-dimensional Mel-generalized cepstral coefficients (MGC) and 511-bin quantized mel scale fundamental frequency (F0) plus one bin for the unvoiced case. These features are extracted from 48-kHz speech waveform using 25-ms window and shifting 5 ms each frame. Linguistic features consist of typical Japanese linguistic information such as phonemes, moras (syllabic unit), part-of-speech tags, interrogative intention, and pitch-accent. The final linguistic features are encoded as a 265-dimensional vector for each frame including duration information extracted from forced-alignment with the acoustic feature sequence, which is obtained using an external systems.

\subsection{Model configurations}
We adopted the same architecture described in our previous publication \cite{luong2018investigating} for the acoustic models. A shallow autoregressive network (SAR) \cite{wang2017autoregressive} is used to model MGC and a deep autoregressive network (DAR) \cite{wang2018autoregressive} is used for quantized mel scale F0. The SAR contains two 512-unit non-linear feedforward layers followed by two 256-unit bi-directional layers, and linear output layer. Similarly, the DAR contains two 512-unit feedforward layers, a 256-unit bi-directional recurrent layer and a 128-unit uni-directional recurrent layer that receives a feedback link from the previously generated samples and a linear layer that maps to the desired output. For the multi-speaker model, a 10-dimensional one-hot vector representing speakers is appended to every frame of the linguistic sequence. The acoustic model is trained using stochastic gradient with the utterance order shuffled to make sure the model learns the optimal representation for all speakers.

A speaker-independent WaveNet vocoder \cite{hayashi2017investigation} was trained using the combined training data of all speakers. 
This model contained 40 dilated layers similar to the original WaveNet  \cite{van2016wavenet}.
It was directly trained using the natural MGC and quantized mel-scale F0s from all the speakers, without speaker one-hot vectors. 
The target waveform had a sampling rate of 16 kHz and was quantized using the 10-bit $\mu$-law standard.

\subsection{Strategies for handling unbalanced corpus}
The main investigation in this paper is which methodology efficiently uses an imbalanced multi-speaker corpus to improve performance for the generated speech of all speakers involved. Multiple strategies are compared in the experiments:
\begin{itemize}
\item \textbf{\texttt{SD}}: 
The conventional speaker-dependent models, each of which is trained using one target speaker's data listed in Table \ref{tab:data}. This is our baseline strategy.
\item \textbf{\texttt{UN}}: A multi-speaker model trained with an under-sampled corpus containing 753$\times$10 utterances. Each speaker contributes 735 utterances to this corpus, where 735 is the number of utterances from speaker XS01, who has the least amount of training data. 
\item \textbf{\texttt{MU}}: The conventional multi-speaker models trained with 
all the data from every speaker, i.e., all 32,076 training utterances from the original corpus.
\item \textbf{\texttt{OV}}: A multi-speaker model trained with an over-sampled corpus. We used all utterances and then sampled more from minority speakers so that each got the same frequency in training. The amount of training data is 8,750$\times$10 utterances.
\item \textbf{\texttt{E1}}, \textbf{\texttt{E2}}, \textbf{\texttt{E3}}: Multi-speaker models trained with resampled corpora. In total, 3,000 utterances are sampled with replication from each speaker. The number of training utterances is 3,000$\times$10, and the number of unique utterances obtained in each sampling session is listed in Table \ref{tab:data}.
\item \textbf{\texttt{EN}}: A non-parametric ensemble model. We simply combined the generated acoustic features obtained from the \texttt{E1}, \texttt{E2} and \texttt{E3} models using the combination functions discussed in Section \ref{subsec:ensemble}.
\end{itemize}

\section{Evaluations}
\label{sec:evaluations}

\subsection{Objective evaluations}

\begin{figure}[t]
  \centering
  \includegraphics[width=\linewidth]{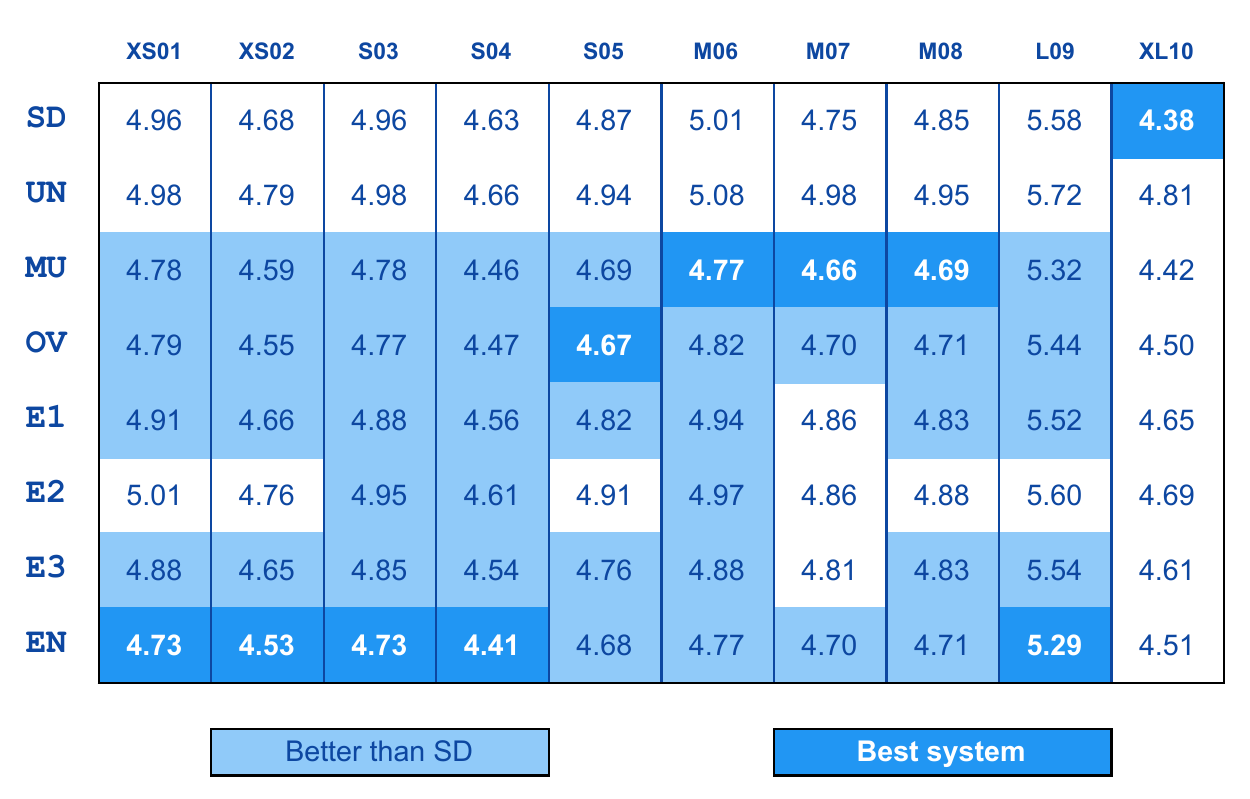}
  \caption{Mel-ceptral distortion (smaller is better).}
  \label{fig:obj-mcd}
  \vspace{-3mm}
\end{figure}

\begin{figure}[t]
  \centering
  \includegraphics[width=\linewidth]{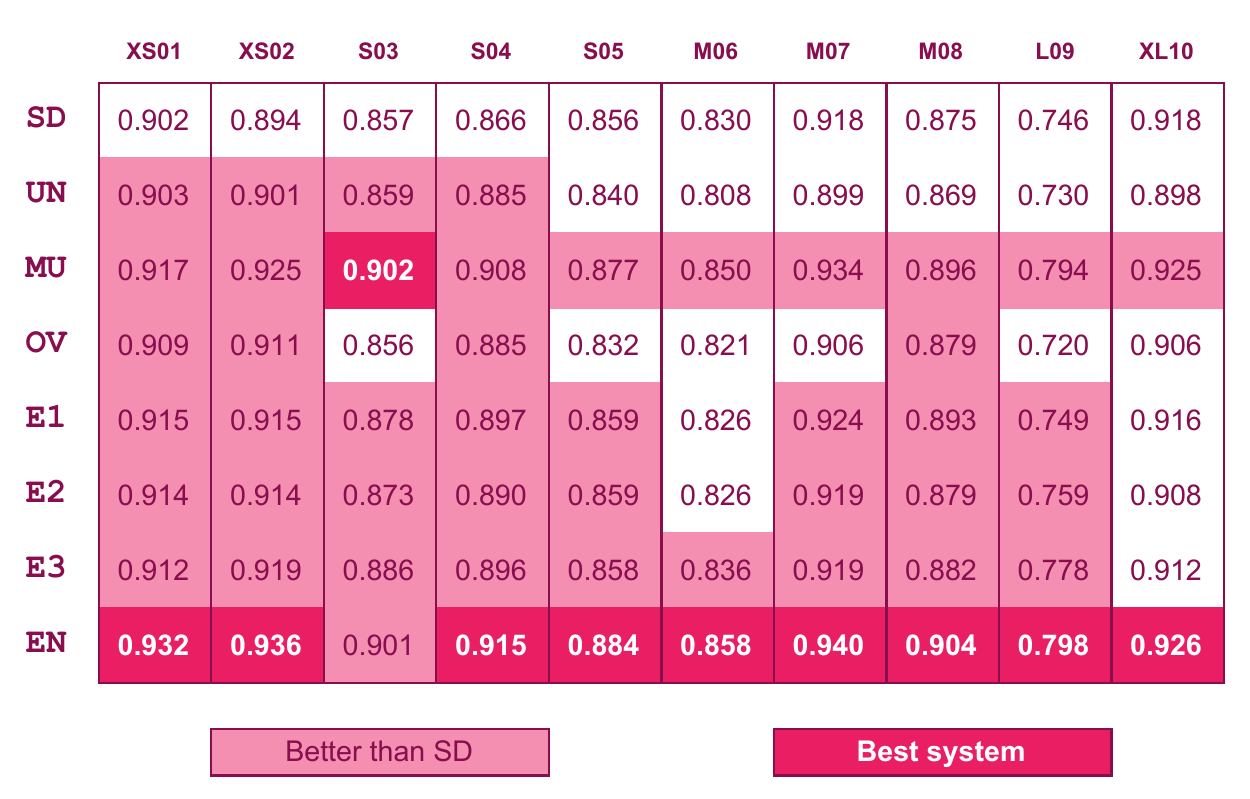}
  \caption{F0 correlation (bigger is better).}
  \label{fig:obj-cor}
  \vspace{-3mm}
\end{figure}

Figure \ref{fig:obj-mcd} shows mel-cepstral distortion between the generated and natural MGC while Fig.\ref{fig:obj-cor} shows correlation between the generated F0 sequence inferred from the quantization output and the natural sequence. These figures show objective results separately for each speaker with color codes indicating the best system as well as the system which is better than the \texttt{SD} baseline. Even though objective evaluations do not directly reflect the quality of synthetic speech perceived by humans, they do demonstrate the potential of the proposed methods.

The under-sampling strategy \texttt{UN} with data pooled from 10 speakers does not seem to have any significant improvement over the baseline \texttt{SD} even for minority speaker XS01, whose entire data is included in \texttt{UN}. This result suggests that a multi-speaker model is not always better than the single speaker model, especially when the amount of pooled data is still limited. The over-sampling strategy \texttt{OV} is better than \texttt{SD} overall, but there is noticeable degradation in the case of majority speakers in terms of the F0 correlation metric. The conventional multi-speaker model \texttt{MU} shows consistent improvements over the baseline \texttt{SD} for most speakers. We conclude that simply pooling the data of all speakers is a reasonable strategy.

The sampling strategies \texttt{E1}, \texttt{E2}, and \texttt{E3} seem to be better than the baseline \texttt{SD} but worse than \texttt{MU}. The performances vary for each session due to the stochastic nature of the sampling method. Surprisingly simply combining the generated features of \texttt{E1}, \texttt{E2}, and \texttt{E3} using the average functions described in Section \ref{subsec:ensemble} produced the a better result than each individual subsystem. In general, the ensemble strategy \texttt{EN} had the best results. Note that the amount of unique utterances from majority speakers (XL10, L09, etc.) used for the ensemble model is significant lower than the \texttt{SD} and \texttt{MU} due to the random sampling artifact, as shown in Table \ref{tab:data}. 

\subsection{Subjective evaluations}

\begin{figure}[t!]
  \centering
  \includegraphics[width=\linewidth]{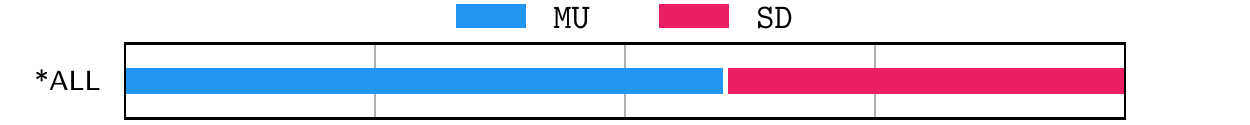}
  \includegraphics[width=\linewidth]{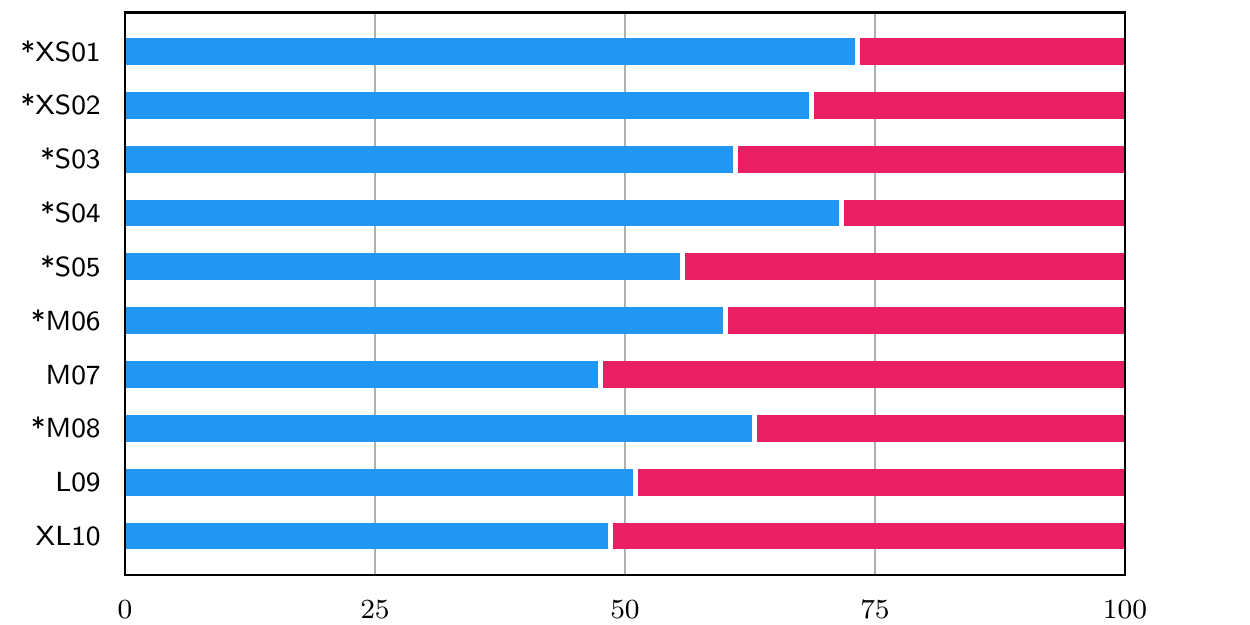}\\
  \vspace{-1mm}
  \texttt{(a) MU-SD} \\
  \vspace{4mm}

  \includegraphics[width=\linewidth]{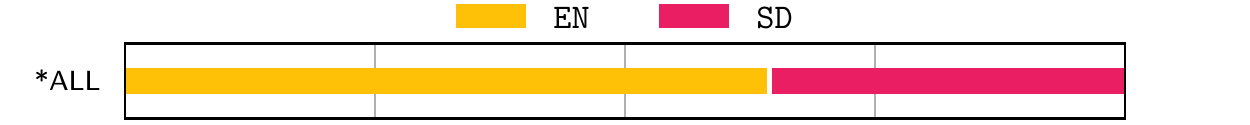}
  \includegraphics[width=\linewidth]{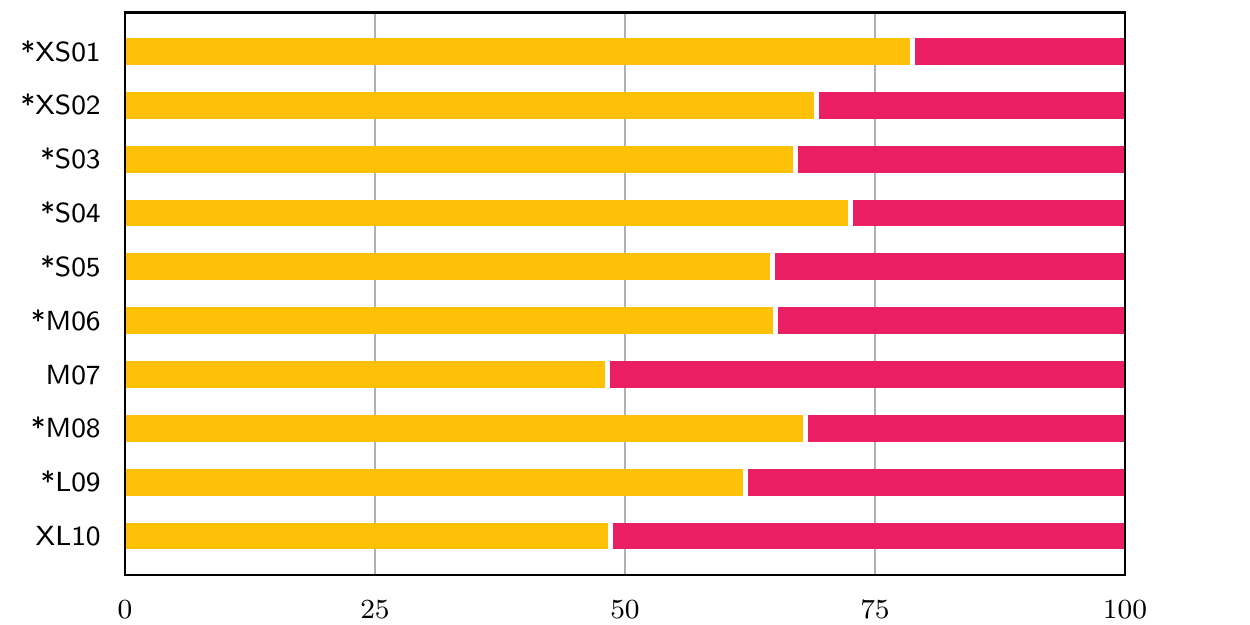}\\
  \vspace{-1mm}
  \texttt{(b) EN-SD} \\
  \vspace{4mm}

  \includegraphics[width=\linewidth]{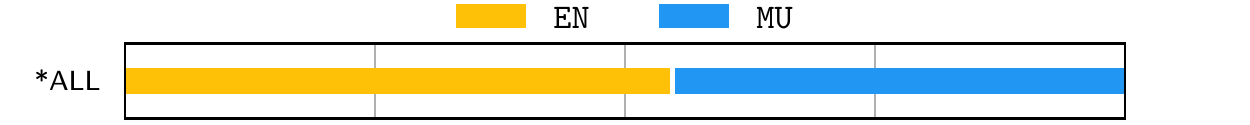}
  \includegraphics[width=\linewidth]{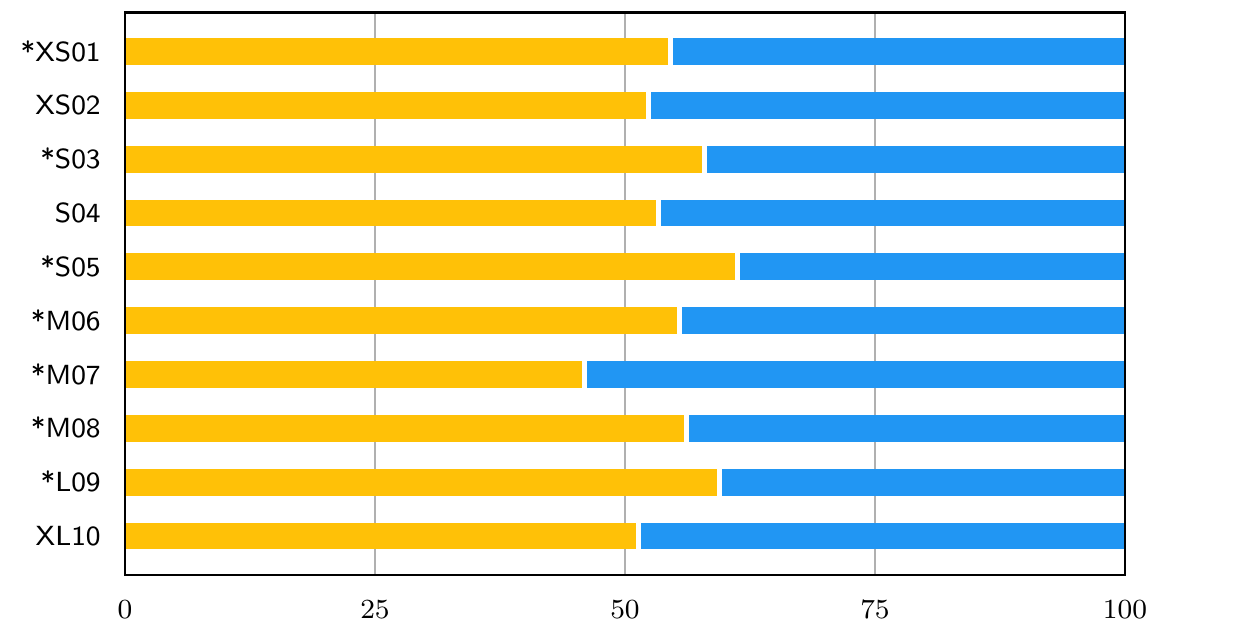}\\
  \vspace{-1mm}
  \texttt{(c) EN-MU} \\
    
  \caption{AB preference test results for TTS samples of three strategies.}
  \label{fig:sub_tts}
  \vspace{-1mm}
\end{figure}

We conducted a subjective listening test with samples synthesized using \texttt{SD}, \texttt{MU} and \texttt{EN} strategies\footnote{Samples are available at \url{https://nii-yamagishilab.github.io/sample-tts-speaker-imbalanced/}}. Recorded speech is not included in our test, but we use WaveNet vocoder to synthesize speech from natural acoustic features as the reference, namely a copy synthesis strategy \texttt{CO}. All samples are normalized using the sv56 program.
Each strategy contains 1,000 utterances, 100 utterances per speaker. We prepared a simple AB preference test in which a participant was asked to answer which sample sounds better between two presented. The presented samples are spoken by the same speaker with the same content and duration but generated from different strategies. We compared four pairs: \texttt{MU-SD}, \texttt{EN-SD}, \texttt{EN-MU} and the anchor test \texttt{EN-CO}. Each session contains one unique sentence from each of the ten target speakers, which make 40 questions in total. The question orders and sample positions are shuffled to prevent cognitive bias. Each paid participant could do ten sessions at most.
We gathered answers from 997 sessions (three are discarded for incompleteness) provided by 175 participants to evaluate performance of the proposed methods. The results are calculated on both a per speaker and per strategy basis.

\begin{figure}[t]
  \centering
  \includegraphics[width=\linewidth]{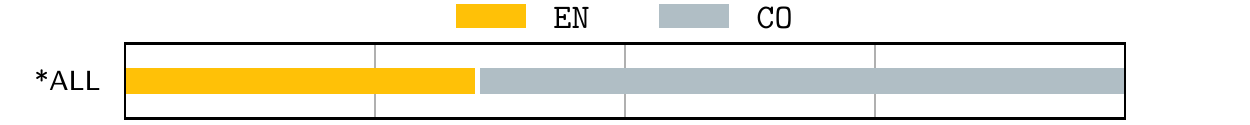}
  \includegraphics[width=\linewidth]{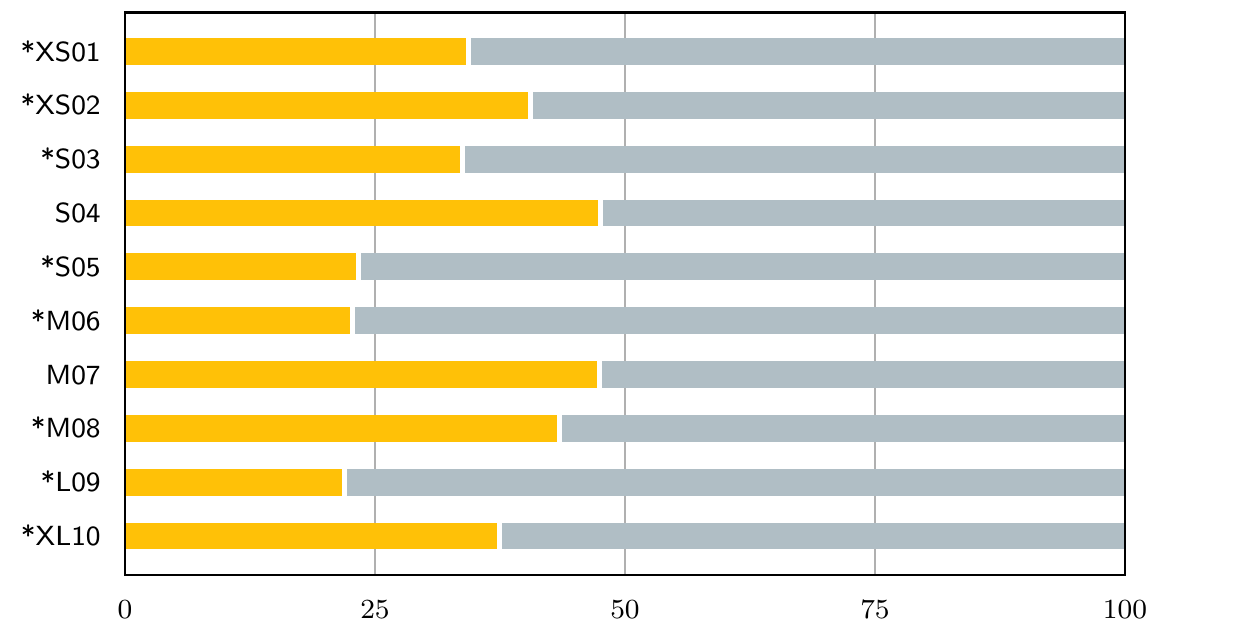}\\

  \caption{Anchor AB preference test results for copy synthesis and ensemble strategy samples.}
  \label{fig:sub_anchor}
  \vspace{-3mm}
\end{figure}

The preference results of the TTS samples are shown in Figure \ref{fig:sub_tts}, 
where (*) indicates systems whose results are statistically significant according to the 95\% confidence level of an exact binomial test.
Between the multi-speaker model and single model, the result is in favor of the \texttt{MU} over the \texttt{SD}, as presented in Fig.\ref{fig:sub_tts}(a). When considering each speaker separately, we can see that speakers with less data benefit the most from the multi-speaker model, while speakers with the most data do not seem to suffer any performance degradation.
A similar pattern can be seen between the ensemble model and the single model (as in Fig.\ref{fig:sub_tts}(b)), with an even stronger improvement observed with the \texttt{EN} strategy.
Figure \ref{fig:sub_tts}(c) shows direct comparisons between the multi-speaker model \texttt{MU} and the ensemble model \texttt{EN}. We obtained statistically significant results favoring \texttt{EN} for many speakers except for M07, who fared best with the \texttt{MU} strategy. The results of speaker XS02, S04 and XL10 while not significant but do seem to favor \texttt{EN} as well. 
To conclude, our proposed ensemble strategy showed significant improvements over the conventional multi-speaker model. The trade-off is the increased number of parameters as well as increased training and inference times due to the fact that multiple models are required.
The anchor test between our proposed strategy \texttt{EN} and the copy synthesis \texttt{CO} is shown in Fig.\ref{fig:sub_anchor}. As expected \texttt{CO} dominated, with statistically significant results for all cases except speakers S04 and M07.

\section{Conclusions}
\label{sec:conclusions}

We investigated the effect of a speaker-imbalanced corpus on the performance of a neural multi-speaker acoustic model. The results showed that simply combining all the available data without any resampling led to a well-rounded performance for all speakers involved. Moreover the multi-speaker model greatly benefited from a simple ensemble setup with just three subsystems sharing the same network structure but trained on different subsets of a corpus obtained through the sampling method. The one disadvantage is that the ensemble setup increases the number of parameters and the inference times. For future work, we plan to distill knowledge from an ensemble teacher network to a singular-structure student to inherit the good performance while avoiding increased parameters and processing times \cite{chebotar2016distilling}. We also intend to introduce diversity to the network structure along with diversity in training data in order to capitalize on the strengths and reduce the weaknesses of different network structures \cite{ren2016ensemble}.


\bibliographystyle{IEEEtran}

\bibliography{main}

\end{document}